# A Hotspot's Better Half:

## Non-Equilibrium Intra-Molecular Strain in Shock Physics

Brenden W. Hamilton[1], Matthew P. Kroonblawd[2], Chunyu Li[1], Alejandro Strachan[1]*

[1]School of Materials Engineering and Birck Nanotechnology Center, Purdue University, West Lafayette, Indiana, 47907 USA
[2]Physical and Life Sciences Directorate, Lawrence Livermore National Laboratory, Livermore, California 94550, United States

**Abstract** Shockwave interactions with material microstructure localizes energy into hotspots, which act as nucleation sites for complex processes such as phase transformations and chemical reactions. To date, hotspots have been described via their temperature fields. Nonreactive, all-atom molecular dynamics simulations of shock-induced pore collapse in a molecular crystal show that more energy is localized as potential energy (PE) than can be inferred from the temperature field and that PE localization persists through thermal diffusion. The origin of the PE hotspot is traced to large intra-molecular strains, storing energy in modes readily available for chemical decomposition.

When materials are subjected to high velocity impacts, a supersonic shockwave is generated that launches a cascade of ultrafast thermal, mechanical, and chemical processes. The extreme temperature and pressure conditions typical of shock loading can lead to the formation of high-pressure phases [1,2] and synthesis of new materials [3,4,5,6], open unexpected chemical pathways for the origins of life [7,8,,9,10], and detonate explosives [11,12,13]. Underpinning our physical understanding of the nucleation of these processes is a widely accepted conceptual framework in which a material's microstructure and defects interact with the shockwave to spatially concentrate energy [12]. In high-energy density materials such as explosives, these nucleation sites, referred to as *hotspots*, are known to govern the shock-to-detonation transition and detonation failure [13]. Hotspots are described by their temperature fields and it is understood that a size-dependent critical temperature needs to be achieved for a hotspot to transition into a deflagration wave [14,15]. In sub-critical hotspots, thermal diffusion dissipates energy before exothermic reactions can lead to self-sustained burn. Developing a predictive understanding of hotspot formation and evolution is critical for improving continuum-level models used to assess intentional and accidental initiation and tailor performance.

Numerous mechanisms are known to generate hotspots during shocks, including pore collapse, jetting, friction, crack propagation, and localized plastic deformation via dislocations or shear bands [16,17,18]. Among these, pore collapse is believed to be the dominant mode for explosive initiation. This was demonstrated through experiments where explosives were rendered non-detonable following a weak shock that collapsed porosity without inducing chemical reactions [19]. Gas gun experiments on gelled nitromethane with porosity controlled through either micro-balloons or low-density silica beads of consistent size showed that the former produces more effective hotspots for initiation, demonstrating the potency of pore collapse in the localization of energy [20]. Regardless of origin, hotspots lead to accelerated chemistry and those above a critical size and temperature can turn into deflagration waves that may ultimately coalesce in a detonation front [21].

Significant efforts have been devoted to understanding how hotspots form and what factors control their criticality. Continuum-level modeling revealed the size dependence of the critical temperature needed to transition to a deflagration [14,22]. Using a model system with a simplified exothermic chemistry, atomistic simulations described the coalescence of deflagration waves into a detonation [23]. Recent reactive molecular dynamics (MD) simulations of solid explosives predicted the transition to deflagration following shock-induced pore collapse [24]. These dynamically formed hotspots are more reactive than counterparts of equal size, temperature, and pressure but created under equilibrium conditions in crystalline or amorphous samples [15]. Dynamic plastic failure can also chemically activate explosive materials and enhance deflagration rates [25]. While experimental observation of shock-induced hotspots remains challenging, recent optical pyrometry measurements have captured peak hotspot temperatures in heterogeneous explosives [26,27,28].

Hotspots are routinely characterized and analyzed by their temperature fields, across atomistic and continuum modelling [14,24,29] and in experiments [30]. That is, hotspots are considered as sites of localized thermal energy. This treatment implicitly assumes that the potential energy (PE) of the material tracks directly with the kinetic energy (KE). It is well understood that the applying external forces to a molecule can accelerate and change chemical reactions through mechanochemistry [31]. This can occur through the force altering the potential energy surface of the molecule, lowering activation barriers [32]. Simple models and theories, such as Bell's Theory [33] and Extended Bell's Theory (EBT) [34], predict that the kinetics depend upon these applied forces. In general, mechanochemistry studies show that stretched bonds typically have shorter lifetimes and that this effect is statistical in nature [35,36,37,38]. Altering the internal conformation of molecules and intra-molecular strain can also result in accelerated chemistry [39,40].

A distinct possibility that we consider here is that pore collapse may lead to conditions that favor a mechanochemical acceleration of reaction kinetics in hotspots. Utilizing large-scale non-reactive MD simulations of pore collapse in 1,3,5-triamino-2,4,6-trinitrobenzene (TATB), we show that energy localization in hotspots is not well characterized by its temperature field alone. We find that the PE of hotspots is not only substantially greater, but also persists for longer times than the temperature field given by KE. This excess PE is found to directly connect with intra-molecular strains, indicating it is localized in modes relevant to chemistry. Neither the initial PE field nor its evolution can be inferred from local temperature fields, yet PE may influence the energy landscape for thermally activated processes. Thus, the overlooked PE component could play a significant role in hotspot dynamics.

**Simulation details**. All MD simulations were conducted using LAMMPS [41], and with interatomic forces described by a widely used [42,43,44,45,47,55,60] all-atom non-reactive TATB force field (TATB FF) [42,43,44]. The TATB FF has been described in detail elsewhere, but we note that its form allows for a rigorous separation of inter- and intra-molecular PE terms. A nearly orthorhombic simulation cell was prepared using the generalized crystal-cutting method [45] starting from the triclinic $P\bar{1}$ TATB crystal structure [46] with lattice parameters determined with the TATB FF at 300 K and 1 atm [47]. The crystal was oriented such that **[100]** was aligned with **x**, **[120]** was nearly parallel to **y**, and the normal to the basal planes $\mathbf{N}_{(001)} = \mathbf{a} \times \mathbf{b}$ was aligned with **z**, the shock direction. A cylindrical pore of radius 40 nm and axis along **x** was created in the geometric center of a cell with (x,y,z) dimensions (3.6 nm, 122.2 nm, 273.1 nm), yielding ~12.3 million atoms. Direct shock simulations were conducted using a reverse-ballistic configuration [48] with an impact speed of 2 km/s, generating a supported shock in target material pre-equilibrated at 300 K. Atomic positions, velocities, and stresses were used to compute molecular properties by local averaging within a 1.5 nm sphere about each molecule. Field properties were obtained through 2D Eulerian binning with (y,z) bin size (2.5 nm, 2.5 nm). Additional simulation and trajectory analysis details can be found in the Supplemental Material (SM) [49,50,51,52].

In explicit shock simulations, the cell size limits the achievable timescales due to rarefaction waves originating from free surfaces. Thus, to extend the simulation time and capture the relaxation mechanisms of the hotspot we developed and employed shock trapping internal boundaries (STIBs), an extension of shock absorbing boundary conditions (SABCs) [53,54]. In this new generalization, fixed boundaries were imposed away from the hotspot to isolate it from the multiple waves reflected from the piston, free surface, and pore collapse event. Implementation details are provided in the SM.

**Energy localization following pore collapse**. As expected for strong shocks with particle velocity 2 km/s [24,29,55], a hydrodynamic collapse of the pore results in significant energy localization and temperature rise. Hotspot formation from pore collapse involves the expansion of the material near the upstream surface of the pore, followed by re-shock against the opposite surface until complete volumetric collapse at time $t_o$. Recompression and shear deformation results in substantial local heating and loss of crystalline order. Long timescale hotspot evolution past $t_o + 27.5$ ps was predicted using STIBs. The resulting temperature field, which is directly proportional to the KE in this classical simulation, exhibits a crescent shape (see Figure 1, bottom row). As expected, the increase in KE plus local compression and disorder also increases local PE. While the PE hotspot eventually settles into a similar crescent shape as the KE, it is greater in magnitude, covers a larger area, and exhibits a different shape during the initial 10 ps following collapse. Interestingly, the PE is mainly stored in intra-molecular terms, indicating significant strain to covalent bonds and molecular shape, see top two rows of Fig. 1. Contrasting this is a largely homogeneous intermolecular PE field, which one might expect to exhibit greater localization from breaking the layered hydrogen-bonding network conjectured to be key to TATB's insensitivity [56,57]. The observed intra-molecular localization implies that much of the excess PE is readily available in modes needed for chemical decomposition, and could plausibly accelerate local kinetics within the hotspot. Importantly, excitation of these intra-molecular modes is nearly instantaneous compared to the lengthy processes of energy up-pumping from lattice modes [58].

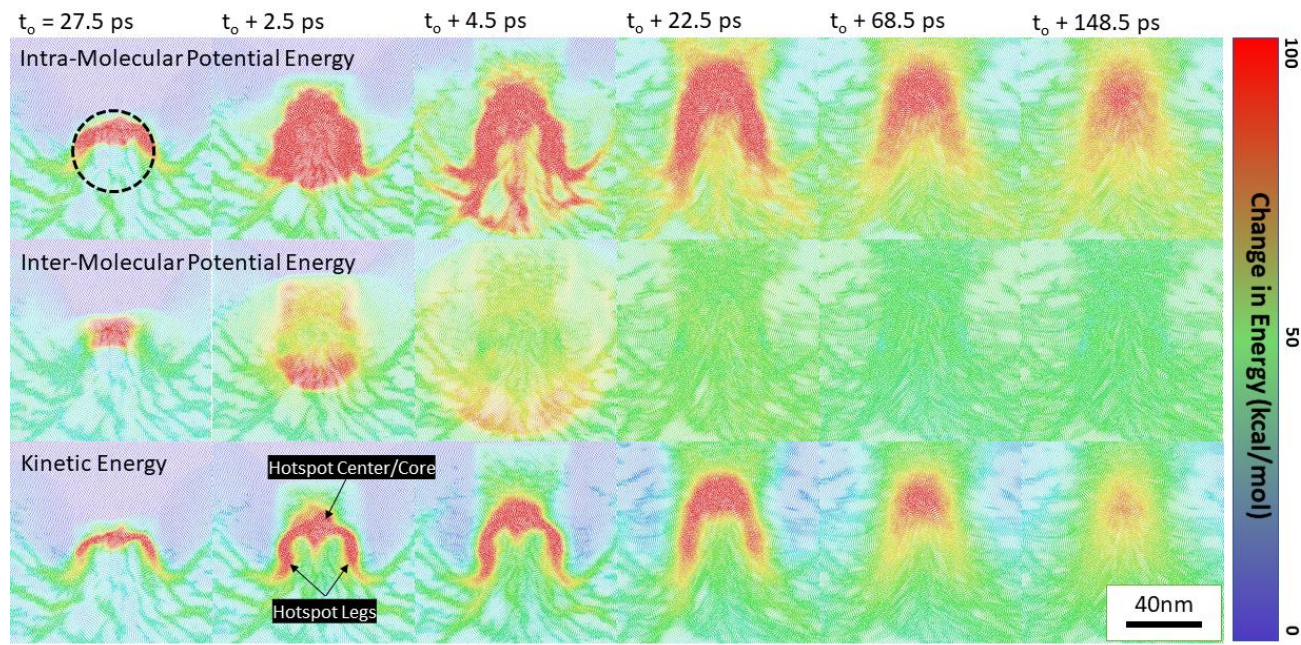

Figure 1: Trajectory snapshots rendered with the OVITO software package [59] showing the temporal evolution of the hotspot in terms of KE (temperature) and PE (separated into intra- and inter-molecular terms). Time $t_o$ represents complete volumetric collapse of the pore. Change in energy is measured with respect to perfect crystal at (300 K, 0 GPa). The black circle in the top left frame represents the initial pore size and location.

The shape of the KE hotspot at early times ($t_o$ + 2.5 ps) reveals the mechanisms behind its origin: a rather circular central region arises from the recompression of expanding material and two *legs* are formed by localized shear deformation. This initial distribution quickly settles into a crescent-like shape that persists for tens of picoseconds. In marked contrast, the early shape of the PE hotspot ($t_o + 2.5$ ps) commands a larger size and persists for longer times within heavily sheared regions ($t_o + 4.5$ ps) before settling to a shape similar to that of the KE component. This decoupling between the KE and PE hotspots indicates a lack of local equilibrium and path-dependent states, discussed in detail below. Reactive MD simulations of shock-induced pore collapse in RDX[24] have shown that the decomposition of a hotspot similar in size to ours requires approximately 40 ps with early reactions seen at the impact plane within ~5 ps. Thus, the timescales in Figure 1 indicate that most of the initial reactions within the hotspot will take place in molecules trapped in excited, highly-strained configurations. This excess PE localization may be the key to understanding the increased reactivity of dynamical hotspots (via mechanochemistry) as compared to those formed thermally under equivalent thermodynamic conditions observed in Ref. 24, as intra-molecular deformations can lead to significant altering of the chemical reactivity [31,39,40].

**Hotspot relaxation characteristics**. To predict the time evolution of KE and PE in the dynamical hotspot, we applied STIBs to extend the simulation time well beyond the passage of the shock. We characterize the intensity and extent of KE and PE hotspots using a function, $A(E)$, that quantifies the area with energy exceeding a value $E$. The functions were obtained from fields computed using Eulerian binning and are shown in Figure 2 for various times. The results confirm that the PE hotspot is larger in terms of area and has a higher specific energy than its KE counterpart and also reveal that it has a substantially longer lifetime. The peak KE value decreases ~1.5 times faster than the peak PE value.

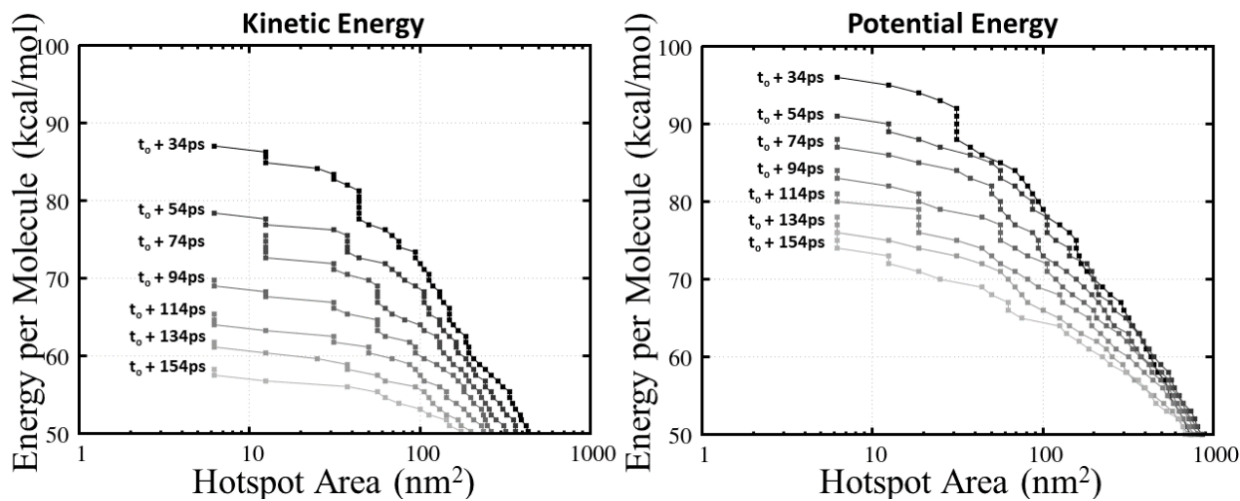

*Figure 2: Hotspot Energy-Size distributions for KE and PE over the extended STIB time regime. Areas are discretized on a 2D Eulerian grid. Energy in both panels is measured with respect to the bulk shocked material, which quantifies the hotspot in terms of excess energy on top of that from hydrodynamic shock compression.*

Since thermally activated chemical processes govern the transition to deflagration, assessing the lifetime of hotspots is key in determining the shock-induced response of explosives. While the relaxation of the hotspot KE (or temperature) can be accurately described via Fourier heat diffusion [60], the mechanisms underlying the evolution of the hotspot PE are not as simple and are governed by fundamentally different physical processes (e.g., conformational changes pressure fluctuations). To study the relaxation of the hotspot's PE, we track the time evolution of the difference in PE and KE of groups of molecules classified based on their degree of internal strain soon after the pore collapse ($t_o + 5$ ps), see Figure 3. Internal strain is quantified by the ratio between the two smallest principal moments of inertia, $I_1$ and $I_2$ of each molecule [61]; this value is near unity for a relaxed TATB molecule as its planar shape leads to moments satisfying $I_1 \approx I_2 < I_3$ (see SM for details). The relatively flat nature of the time histories in Fig. 3 indicates negligible relaxation of the highly strained molecules beyond the thermal component that tracks the reduction in KE. That is, a significant portion of the PE rise is latent and undergoes minimal relaxation (the molecules remain highly distorted) on time scales long enough for the KE localization to be completely dissipated. Any significant structural relaxations of the highly deformed molecules would result in a large negative slope for the blue and brown curves in Figure 3. Individual histories of PE and KE are available in SM.

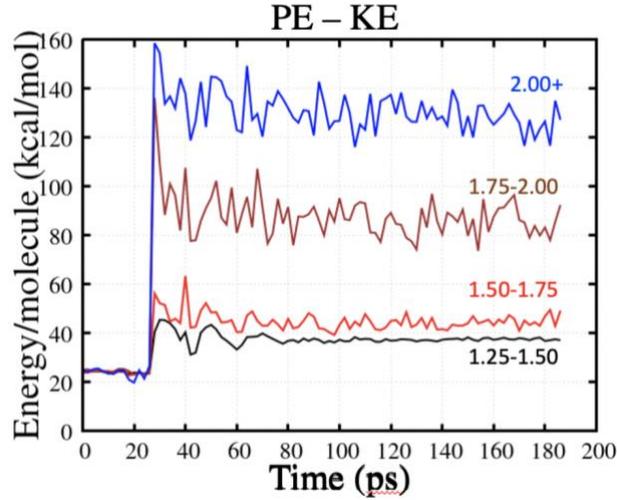

*Figure 3: Time history of the difference between the change in PE and change in KE for all molecules within a 25nm cylindrical radius of the hotspot center. Curves correspond to averages over molecules grouped by the ratio of their 1st and 2nd principal moments of inertia $I_2/I_1$, which has a value of 1.0 for a disk. Comparatively fewer molecules have large $I_2/I_1$, leading to larger fluctuations in the blue and brown curves.*

**Spatial localization of potential energy.** Having established that local temperature does not fully determine the thermodynamic state of the system, we now seek a more complete characterization of PE localization. Figure 4(a) shows local PE of each bin as a function of its temperature at $t_o + 2.5$ ps. Curves for PE vs. $T$ in crystalline and amorphous samples at the shock pressure ($P = 22$ GPa) are included for comparison. This reveals a broad distribution of local states following pore collapse that is decidedly unlike either crystal or amorphous samples. The lack of a one-to-one PE-T relationship confirms that the hotspot state is not uniquely characterized by temperature. Furthermore, even a two-phase crystal/amorphous description of the system does not fully capture the underlying complexity found in local states at these early stages. Local states were divided into arbitrary categories (marked by the different colors) and mapped into real space in Fig. 4(b). Besides the unshocked material (black), the elastic precursor following the shock front (purple), and plastic (orange) shocked crystal regions, we identified two distinct zones associated with the hotspot: A core (lime) surrounded by a halo (forest green). The local PE states within the hotspot core and halo regions span a wide range. The core states lie primarily above the amorphous baseline at early times due to high pressure from pore collapse re-shock. The halo ranges from above the amorphous curve down to crystal-like values. Note that the maximum temperature is lower than the predicted melting point for TATB ($T_m = 3400$ K at 22 GPa) [62], indicating that the core and halo are highly plastically deformed solids that may explain locked-in intra-molecular strain. Plots similar to 4(a) at later times are available in SM.

The use of a non-reactive force field enables us to separate the thermo-mechanical processes associated with hotspots from chemistry. Future reactive simulation can be expected to shed light into the acceleration of chemistry by the mechanical excitation identified here. We note that if chemistry is accelerated by the PE component of the hotspot, comparison with earlier work suggests these reaction timescales will be longer than that of hotspot formation and more similar to the timescale of PE relaxation.

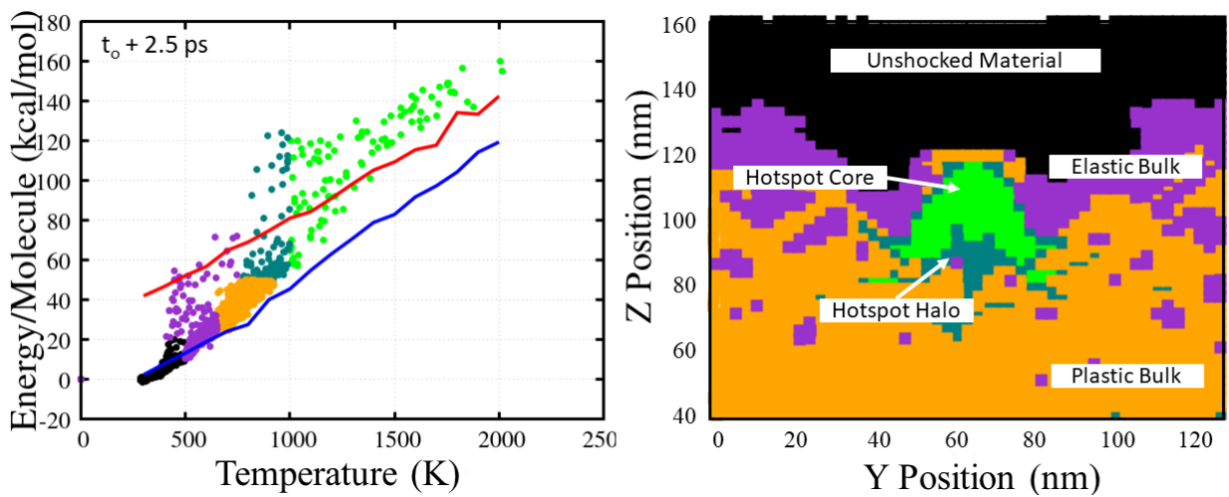

*Figure 4: (a) Plot of PE as a function of temperature (points) compared to isobars for a crystal (blue line) and amorphous (red line) system. (b) System divided into zones based on PE and T showing parts that are unshocked (black, PE < -20 kcal/mol, T < 500 K), elastically compressed crystal (purple, 500 < T < 650 K), plastically deformed crystal (orange, -20 < PE < 10 kcal/mol, 650 < T < 1000 K), the hotspot halo region (forest green, PE > 10 kcal/mol, 750 < T < 1000 K), and the hotspot core (lime, T > 1000 K).*

State-of-the-art continuum-based models describe hotspot kinetics and chemistry in terms of temperature rise [63,64], using chemistry models that, at most, distinguish between solid and liquid phases [14] but do not account for disordered solids or the scatter in local PE revealed by our simulations. Furthermore, these models describe energy transport in terms of thermal diffusivity if not outright neglected via an adiabatic approximation. The rise in PE adds another layer to the competition between exothermic reactions and energy dissipation that controls hotspot criticality, since the localized PE may alter the material reactivity, and its dissipation is governed by physics very different from thermal diffusion. Persistence of highly deformed molecules leads to an acceleration of local reaction kinetics in a variety of situations [10,25,40,65,66,67]. This latent PE persists over timescales similar to the initial exothermic chemistry predicted in reactive MD pore collapse simulations [24,68], which could help to explain why dynamically formed hotspots are more reactive than those in pure crystalline or amorphous samples [15].

In summary, we show through large-scale MD simulations that the PE states in hotspots resulting from shock-induced pore collapse in molecular crystals exhibit a wider spatial extent and have higher energy density with a more complex distribution than can be inferred from the temperature field. These excited PE states arise due to significant distortions of molecular geometry that persist well beyond the dissipation of the temperature field. Our results show that persistent molecular deformations necessary for mechanochemistry arise in shocked porous molecular solids and can be measured in simulations via an easily computed metric, the intra-molecular PE. This offers a plausible and testable hypothesis to explain prior puzzling MD results in which mechanically formed hotspots were found to be more reactive than thermally formed ones. The large differences between the temperature and PE fields identified here motivates a renewed analysis of hotspot dynamics, including attendant plastic work, phase transformations, and chemical kinetics, in a way that considers path dependent mechanochemistry arising in highly strained regions. More generally, our simulations highlight a neglected physical aspect of the early stages of hotspot formation and evolution that may offer a route to improve multiphysics models of shock initiation and detonation.

**Acknowledgments**

This work was supported by the Laboratory Directed Research and Development Program at Lawrence Livermore National Laboratory, project 18-SI-004 with Lara Leininger as P.I. Partial support was received from the US Office of Naval Research, Multidisciplinary University Research Initiatives (MURI) Program, Contract: N00014-16-1-2557. Program managers: Clifford Bedford and Kenny Lipkowitz. Simulations were made possible by computing time granted to MPK through the LLNL Computing Grand Challenge, which is gratefully acknowledged. This work was performed under the auspices of the U.S. Department of Energy by Lawrence Livermore

# Supplemental Material

# A Hotspot's Better Half:
## Non-Equilibrium Intra-Molecular Strain in Shock Physics

Brenden W. Hamilton[1], Matthew P. Kroonblawd[2], Chunyu Li[1], Alejandro Strachan[1]*

[1]School of Materials Engineering and Birck Nanotechnology Center, Purdue University, West Lafayette, Indiana, 47907 USA
[2]Physical and Life Sciences Directorate, Lawrence Livermore National Laboratory, Livermore, California 94550, United State

## A. TATB Force Field

All classical molecular dynamics (MD) simulations were performed using a validated variant of the rigid-bond non-reactive non-polarizable force field (FF) for 1,3,5-triamino-2,4,6-trinitrobenzene (TATB) originally developed by Bedrov et al. [1]. The version of the FF used here includes tailored harmonic bond stretching and angle bending terms to yield (nearly) fully flexible molecules [2], RATTLE constraints to fix the N-H bond stretching motions to their equilibrium values [3], and an intramolecular O-H repulsion term [4] that is implemented as a bonded interaction. Using RATTLE constraints enables a longer time step and improves the description of the heat capacity. The covalent bonds, three-center angle bends, and improper dihedrals are modeled using harmonic functions and the proper dihedrals are modeled using cosine series. Non-bonded interactions are modeled using the Buckingham potential (exp-6) with short-ranged Lennard-Jones-style $r^{-12}$ repulsion terms and electrostatic interactions between fixed partial charges located on the nuclei. All intramolecular non-bonded interactions are excluded by design, which allows for rigorous separation of inter- and intra-molecular potential energy terms. Buckingham/Lennard-Jones potential interactions were evaluated in real space within an 11 Å cutoff and electrostatic interactions were evaluated using the short-ranged Wolf potential [5] with a damping parameter of 0.2 Å$^{-1}$ and an 11 Å cutoff.

## B. Classical Molecular Dynamics Methods

Prior to any trajectory integration, we generated free surfaces normal to **z** (the shock direction) that lift the periodicity in that direction by adding a 5 nm region of vacuum padding to prevent self-interactions. A thermal configuration was obtained at 300 K from a 25 ps canonical (*NVT*) trajectory integrated using a Nose-Hoover-style thermostat [6,7] and a 0.5 fs timestep. During the first 2.5 ps, we reselected atomic velocities every 0.5 ps from a Maxwell distribution and rescaled the velocities to the target temperature every 0.05 ps to attenuate breathing modes induced by exposing free surfaces. This configuration was used as the starting point for a reverse ballistic shock simulation that was integrated using micro-canonical (*NVE*) equations of motion and a 0.2 fs timestep. In the reverse ballistic setup, a piston velocity $\mathbf{u}_P = (-2 \text{ km/s})\hat{\mathbf{z}}$ was added to the thermal velocities of atoms in a flexible sample leading to impact on the rigid piston that generated a shock front traveling through the sample in the opposite direction to $\mathbf{u}_P$. Molecules with center of mass positions with $z \leq 1.5$ nm were held fixed throughout the shock simulation to simulate a rigid and infinitely massive piston.

## C. Trajectory Analysis

Most of our trajectory analysis was performed within a molecular framework starting from per-atom positions $\mathbf{r}_i$, velocities $\mathbf{v}_i$, and stress tensors $\mathbf{s}_i$. From the per-atom quantities we compute the molecular center of mass (CM) positions, velocities, and stresses as

$$\mathbf{R} = \frac{1}{M}\sum m_i \mathbf{r}_i \quad \text{(SM-1)}$$

and

$$\mathbf{V} = \frac{1}{M}\sum m_i \mathbf{v}_i \quad \text{(SM-2)}$$

and

$$\mathbf{S} = \sum \mathbf{s}_i \qquad \text{(SM-3)}$$

where $M$ is the mass of a TATB molecule, $m_i$ is the mass of atom $i$, and the sum runs over all 24 atoms in the molecule. From these quantities, we compute the total molecular kinetic energy $KE_{tot}$, and the separate contributions from the molecular translational $KE_{trans}$ and roto-librational and vibrational $KE_{ro\text{-}vib}$ degrees of freedom as

$$KE_{tot} = \sum \frac{1}{2} m_i \mathbf{v}_i \cdot \mathbf{v}_i \qquad \text{(SM-4)}$$

and

$$KE_{trans} = \frac{1}{2} M \mathbf{V} \cdot \mathbf{V} \qquad \text{(SM-5)}$$

and

$$KE_{ro-vib} = KE_{tot} - KE_{trans} \qquad \text{(SM-6)}$$

The ro-vib kinetic energies $KE_{ro\text{-}vib}$ are interpreted as the molecular temperature $T$ and are scaled to Kelvin units through

$$KE_{ro-vib} = \frac{63}{2} k_B T \qquad \text{(SM-7)}$$

where $k_B$ is the Boltzmann constant and the factor of 63 arises from the 3 roto-librational and 60 unconstrained vibrational degrees of freedom in the TATB molecule. It should be noted that LAMMPS-computed per-atom stresses $\mathbf{s}_i$ are in units of stress*volume and must be normalized by a meaningful volume to yield the actual local stress. We obtained molecular stresses $\sigma$ in stress units by summing all per-molecule $\mathbf{S}$ within a sphere of radius 1.5 nm about a given molecule and taking the volume of that sphere as the normalizing volume.

Distortions to molecular geometry were quantified through the molecular inertia tensor in the principal rotational frame. First, the inertia tensor $\mathbf{I}$ for a molecule was computed as a sum over point particles in a Cartesian lab frame in which the atomic coordinates of the molecule were translated to place the CM at the origin. Diagonalizing $\mathbf{I}$ yields the principal moments of inertia and axes for the principal rotational frame as the eigenvalues and eigenvectors, respectively. Because the TATB molecule is essentially a $D_{3h}$-symmetric disk at equilibrium, the two smallest principal moments $I_1$ and $I_2$ are nearly equivalent and have axes that lie within the plane of the molecule. The third moment and axis correspond to rotation about a vector that is normal to the plane of the molecule. Performing this analysis for the entire trajectory involves $O(10^8)$ diagonalizations of a real symmetric 3x3 matrix, so we used the efficient *QL* algorithm implemented in Fortran by Kopp [8].

### D. Shock Trapping Internal Boundaries

In previous works using Shock Absorbing Boundary Conditions (SABCs) [9] [10], a planar wave is either absorbed by a rigid body at a free surface or annihilated by a duplicate wave at a periodic boundary. Both these approaches rely on a wave front that is moderately planar and specific, premeditated boundary conditions. In pore collapse simulations, and other simulations of microstructure interactions with shock waves, multiple reflections and re-shock waves are created that must be attenuated. At least three additional waves arise in simulations involving collapse of a single cylindrical pore. While absorbing these multiple waves (all moving at different speeds) would be difficult, we aim to simply trap them to prevent their interaction with an area of interest, namely the hotspot. Shock Trapping Internal Boundaries (STIBs) are a generalization of SABCs that we develop here to soften the above requirements and at the same time increase computational efficiency for problems involving non-uniformly shaped regions of interest.

We apply a set of STIBs at a point in time within the simulation that all shockwaves are sufficiently far from the area of interest, in this case a dynamically formed hotspot. Unlike SABCs which are fixed immobile regions that *encompass* a wave front, the STIBs are comparatively small immobile material regions that are imposed *between* the shockwaves and hotspot (see Fig. SM-1). These immobile regions must be wider than the cutoff distance for force interactions between system particles (here, 11 Å) and ideally should be of similar density to the bulk shocked material. In LAMMPS, particles in STIBs are held immobile by leaving them *unassigned* to a time-integration fix. The area of interest and surrounding material is contained between the STIBs and is treated as flexible. With this set of constraints, the material around the hotspot evolves according to interactions between the flexible particles in an approximately uniform field that mimics the bulk that is defined by explicitly computed interactions with the immobile STIB particles. Material on the exterior of the STIBs does not interact with the interior and can be deleted to significantly increase computational efficiency. In the present case, we define the two STIB regions as 30 Å thick slabs that do not cut molecules, which forms a rectangular region containing the hotspot.

The main advantage of STIBs over SABCs is that they can be made in arbitrary shapes with the only constraint being that the region of interest be entirely enclosed (considering periodic boundaries). For instance, a suitable alternative in the present case would be a single STIB in the shape of a cylindrical shell surrounding the hotspot. Parallelepiped-shaped STIBs can efficiently trap other dynamically formed defect regions, such as individual shear bands.

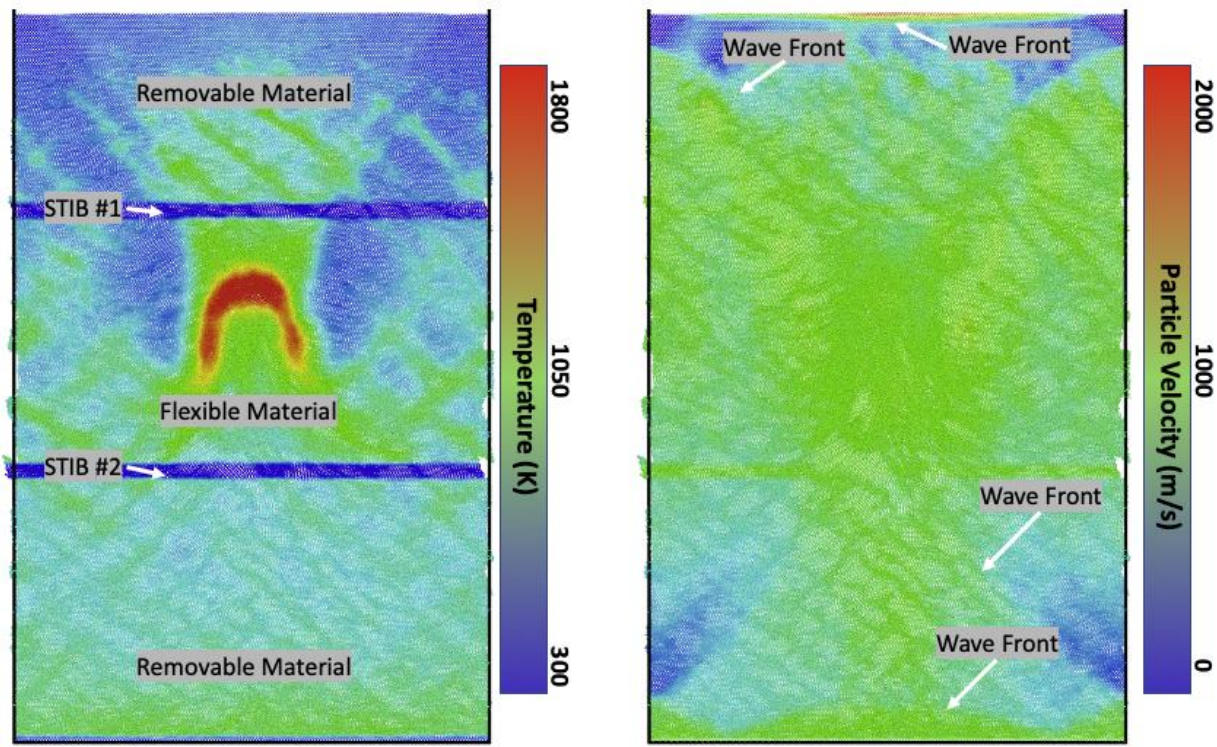

*Figure SM-1: Rectangular STIBs to isolate a dynamically formed hotspot following pore collapse. Over 60% of the material can be safely removed. Black lines denote periodic boundaries of the simulation cell.*

### E.     Full Distribution of Molecular Energies

Figure SM-2 shows the full distribution of all molecule PE values as a function of $T$ for the inset region at $t_o + 22.5$ ps. This scatter-plot representation highlights the full range of available states that are averaged over in the 2D Eulerian binning scheme used for Figure 4 in the main article. At a given $T$, the width of the molecular PE distribution is upwards of 100 to 150 kcal/mol (not including outliers/tails of distributions).

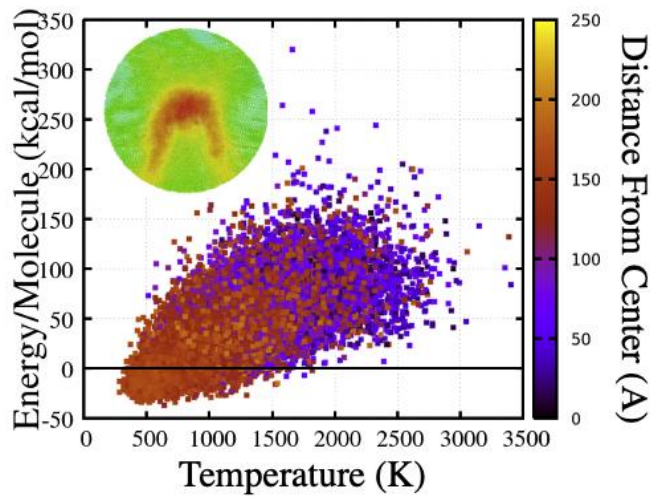

*Figure SM-2: Scatter plot of molecular PE(T) values within a cylinder of radius of 25 nm (see inset). PE values are measured with respect to the average PE of the bulk shocked material. Points are colored by their radial distance from the center of the hotspot.*

### F. Local Binned Energies over Time

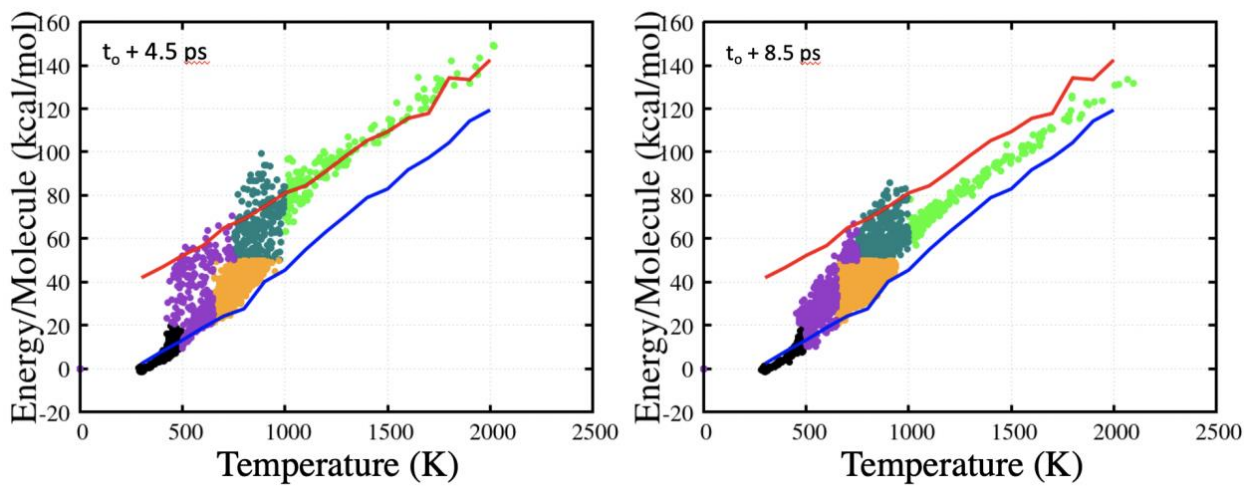

*Figure SM-3: PE-T plots analogous to Figure 4(a) in the main article taken at later times in the simulation. After local equilibrium is reached at $t_o + 8.5$ ps, the qualitative features (e.g., shape and width) of the distribution remain relatively static, but decrease in magnitude due to thermal transport.*

### G. Local Decay of Energy over Time

Figure SM-4 shows time histories for PE and KE from which we compute the differences in PE and KE plotted in Figure 3 of the main article. Each line represents a group of molecules binned by the ratio of their two smallest principal moments of inertia, $I_2/I_1$ (see last paragraph of Section SM-C). The differences between PE and KE shown in Figure 3 of the main article reveal that the slopes of PE-KE are largely constant for each bin. At the same time, the PE and KE values individually decrease over time at approximately the same rate (for each bin) due to thermal

conduction. One would expect that the PE would decrease at a constant temperature during a first-order phase transformation, which is not the case here. This may indicate that the phenomenology of material amorphization is not well described by a first-order phase transformation process.

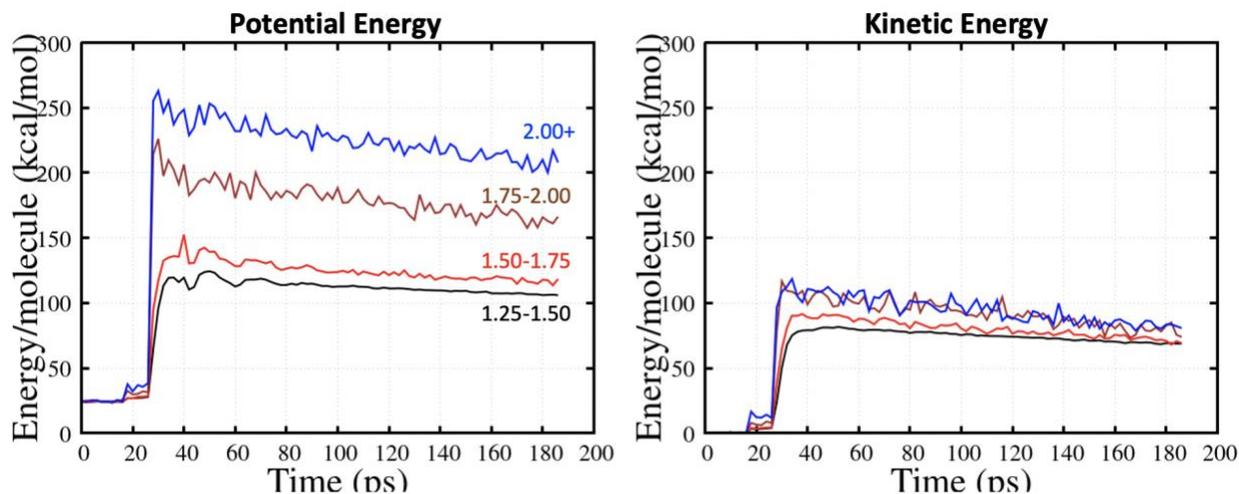

*Figure SM-4: Time histories of PE and KE in hotspot material for different binning ranges of the ratio in the two smallest principal moments of inertia, $I_2/I_1$.*

**Acknowledgements**